\documentclass[aip,apl,a4paper,amsmath,amssymb,floatfix,reprint,superscriptaddress]{revtex4-2}
\usepackage{mathptmx}
\usepackage{float}
\usepackage[scaled=0.9]{helvet}
\usepackage[utf8]{inputenc}
\usepackage{graphicx}
\usepackage[pdftex,dvipsnames]{xcolor}
\usepackage{upgreek}
\usepackage{bm}
\usepackage{textcomp}
\usepackage[
,textwidth=17.5cm
,textheight=23.5cm
,verbose
,dvips
]{geometry}
\usepackage{xspace}
\usepackage{xargs}
\usepackage{verbatim}

\usepackage[normalem]{ulem}

\usepackage[pdftex]{hyperref}  
\hypersetup{pdftitle={},colorlinks,citecolor=blue}

\pdfoutput=1
\begin{document}


\title[]{Transition from edge- to bulk-currents in the quantum Hall regime}

\author{S. Sirt}\email{sirt@pdi-berlin.de}\affiliation{Paul-Drude-Institut f\"ur Festk\"orperelektronik, Leibniz-Institut im Forschungsverbund Berlin e.V., Hausvogteiplatz 5-7, 10117 Berlin, Germany}
\author{V.\,Y. Umansky}\affiliation{Weizmann Institute of Science, 234 Herzl Street, Rehovot 76100, Israel}
\author{A. Siddiki}\affiliation{Istanbul Atlas University, Hamidiye, Anadolu Cd. no:40, 34408, Istanbul, Turkey}
\author{S. Ludwig}\email{ludwig@pdi-berlin.de}\affiliation{Paul-Drude-Institut f\"ur Festk\"orperelektronik, Leibniz-Institut im Forschungsverbund Berlin e.V., Hausvogteiplatz 5-7, 10117 Berlin, Germany}

\begin{abstract}
The integer quantum Hall effect can be observed in a two-dimensional conductor penetrated by a perpendicular magnetic field and with edges connecting the current carrying contacts. Its signature is a state of quantized Hall and simultaneously vanishing longitudinal resistances. A widely accepted model is the Landauer-B\"uttiker picture, which assumes an incompressible, i.e., electrically insulating bulk state surrounded by current carrying one-dimensional edge channels. This single-particle model is challenged by the screening theory. It derives, that electron-electron interaction leads to a fragmentation of the Hall bar into compressible and incompressible strips, where the current flows inside the incompressible strips. Because the latter gradually shift from the sample edges into the bulk as the magnetic field is increased, it suggests a transition from edge- to bulk-current. We present a direct experimental proof of this transition. Our results support the screening theory.
\end{abstract}


\maketitle


Since its dicovery in 1980 \cite{vonKlitzing1980}, the integer quantum Hall effect (QHE) has been a subject of fundamental research and today provides the electrical resistance standard \cite{si-brochure2019,Delahaye2003}. The QHE is a manifestation of the Landau quantization of the charge carriers density of states (DOS) when exposed to a magnetic field. Its main features are extended plateaus of the Hall resistance as a function of the magnetic field at quantized values $R_\nu=R_\mathrm{K}/\nu$, where $R_\mathrm{K}={h}/{e^2}$ is the von Klit\-zing constant and the filling factor $\nu$, measuring the fraction of occupied Landau levels (LLs){, is integer}. Despite the relative maturity of the QHE, its microscopic nature has been controversially discussed since its discovery \cite{Halperin1982,Buettiker1988,Fontein1991,Fontein1992,Klass1991,Chklovskii1992,Geller1994,Zhitenev1994,Yahel1996,Mani1996,Mani1997,Mani1997-2,Yacoby1999,McCormick1999,Weitz2000,Komiyama2006,Dohi2007,Gerhardts2008,Siddiki2009,Siddiki2010,Weis2011,Kendirlik2013,Kendirlik2017,Gerhardts2020}. It goes without saying that the microscopic details are important for both, our fundamental understanding of the QHE and our ability to optimize the accuracy of its metrological application. In particular, the question where the current flows and whether it is coherent is central for possible quantum technology applications, because the phase accumulated by a charge carrier crucially depends on its path way \cite{Ji2003,Goldman1995}. Further, our model of the integer QHE serves as a basis for understanding its relatives, such as the fractional, spin or anomalous  QHEs.

The Landauer-Büttiker (LB) picture \cite{Buettiker1986,Buettiker1988} considers non-interacting electrons in equilibrium. It predicts for the plateaus chiral current flow through $\nu$ quasi one-dimensional (1D) edge channels. The bulk of the sample is assumed to be insulating, caused by the Landau gap of the DOS, where the gap is broadened by localization induced by the magnetic field in a disorder potential \cite{Buettiker1988}. Assuming complete suppression of back-scattering inside the chiral edge channels, one can follow that each (spin-resolved) edge channel would carry the quantized conductance of a perfect 1D channel, $R_\mathrm{K}^{-1}$.

\begin{figure}[t]
\centerline{\includegraphics[width=1\columnwidth]{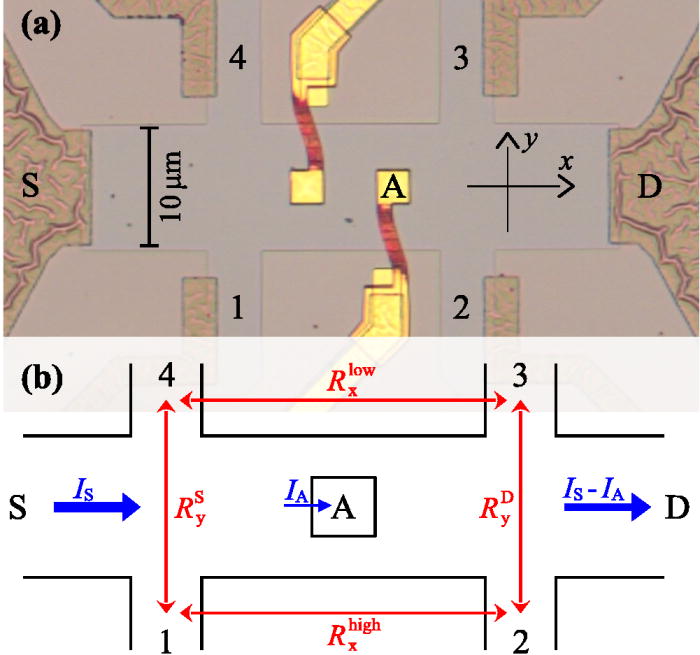}}
\caption{(a) Microscopic photograph of the sample. The mesa (gray color) shapes the $10\,\upmu$m wide Hall bar containing the 2DES. Where it appears brown and smooth, the surface is etched to a depth of $\simeq240$\,nm. Darker brown and rougher regions are the ohmic contacts S, D, 1--\,4. Bright yellow regions are gold gates with air bridges (red) connecting to two inner ohmic contacts (hidden below the squared  $2.5\,\upmu$m wide gold regions). 
(b) Sketch showing contacts, imposed current contributions and resistance matrix elements; voltage probes 1--\,4 are floated.
}
\label{fig:sample}
\end{figure}
The screening theory goes beyond the LB picture by deriving the local electrostatic potential landscape in a self-consistent calculation, while taking into account the direct Coulomb interaction of the charge carriers \cite{Chklovskii1992,Chklovskii1993a,Fogler1994,Lier1994}. This is usually done within the Thomas-Fermi approximation while the quantum nature of the electrons can be included within a semi-classical mean field approach \cite{Siddiki2003,Siddiki2004,Gerhardts2008}. For the case of equilibrium (zero imposed current), the electrostatics predicted by the screening theory was recently confirmed (for $\nu\le3$) by self-consistently solving the full quantum-electrostatic problem  \cite{Armagnat2020}.

According to the screening theory, for a large perpendicular magnetic field, the Landau quantization gives rise to a local bandgap between either completely filled or empty LLs, such that incompressible strips (ICS) emerge, in which electron scattering is completely absent (at low enough temperature). The ICSs separate  compressible regions, inside which a LL is partly filled, so that electrons can scatter. The compressible regions are perfectly screened with the partly filled LL pinned to the local chemical potential. Therefore, both, the energies of the LLs and the chemical potential vary inside the ICSs only.

An imposed current modifies the electrostatics including the potential gradients, which are non-zero only within the ICSs. Being proportional to the local potential gradient, the current density, too, is non-zero only inside the ICSs, while the compressible regions remain free of current \cite{Chklovskii1993a,Guven2003,Gerhardts2008}. This conclusion becomes really evident for the case of a sizable imposed current, i.e., far from equilibrium \footnote{We note, that we disagree with a recent suggestion of the current flowing along the interfaces between ICSs and compressible regions, based on applying the continuity equation at equilibrium \cite{Armagnat2020}. However, even if this suggestion was correct, it would not change the interpretation of our experiment, that we find a transition from edge- to bulk-current along the plateaus of quantized Hall resistance.}. Inside the ICSs the Landau gap suppresses scattering, such that the longitudinal resistance vanishes along the Hall bar, while the dissipation happens in the ohmic contacts. Integrating the current density over the ICSs then gives rise to the observed quantized Hall resistance $R_\nu=R_\mathrm{K}/\nu$. Only, once the imposed current exceeds the breakdown limit of the QHE, the ICSs vanish and the transport becomes diffusive \cite{Kaya2000,Yu2018,Haremski2020}.

In contrast to LB edge channels, the width of an ICS grows as the magnetic field $B$ is increased along the quantized Hall plateau. While growing wider, the ICSs gradually move from the edge towards the center of the Hall bar, where they combine into a single ICS. The exact geometry of the ICSs depends on the local electrostatic potential including the confinement, the Hall potential and the screening via electron-electron interaction. A disorder potential, which is neglected in most calculations, would counteract the predicted asymmetric development of the ICSs along a quantized plateau. However, in contrast to the LB picture, the screening theory predicts plateaus with finite widths even for zero disorder \cite{Siddiki2007}. In order to experimentally test the screening theory, we keep the influence of disorder small by using a relatively narrow high mobility sample, such that its mean-free-path exceeds the Hall bar width.

An increasing number of experiments support the screening theory and thereby provide indirect evidence for the existence of ICSs \cite{Fontein1991,Zhitenev1994,McCormick1999,Weitz2000,Horas2008,Siddiki2009,Siddiki2010,Weis2011,Kendirlik2013,Kendirlik2017} or bulk-currents \cite{Ferguson2023} in the regions of quantized plateaus. We go one step further by performing a direct measurement of bulk-current while the Hall resistance is quantized. To achieve this, we use an inner ohmic contact placed near the center of our  high-mobility Hall bar. Our result contradicts the LB picture but supports the screening theory.


%
\begin{figure}[b]
\centerline{\includegraphics[width=1\columnwidth]{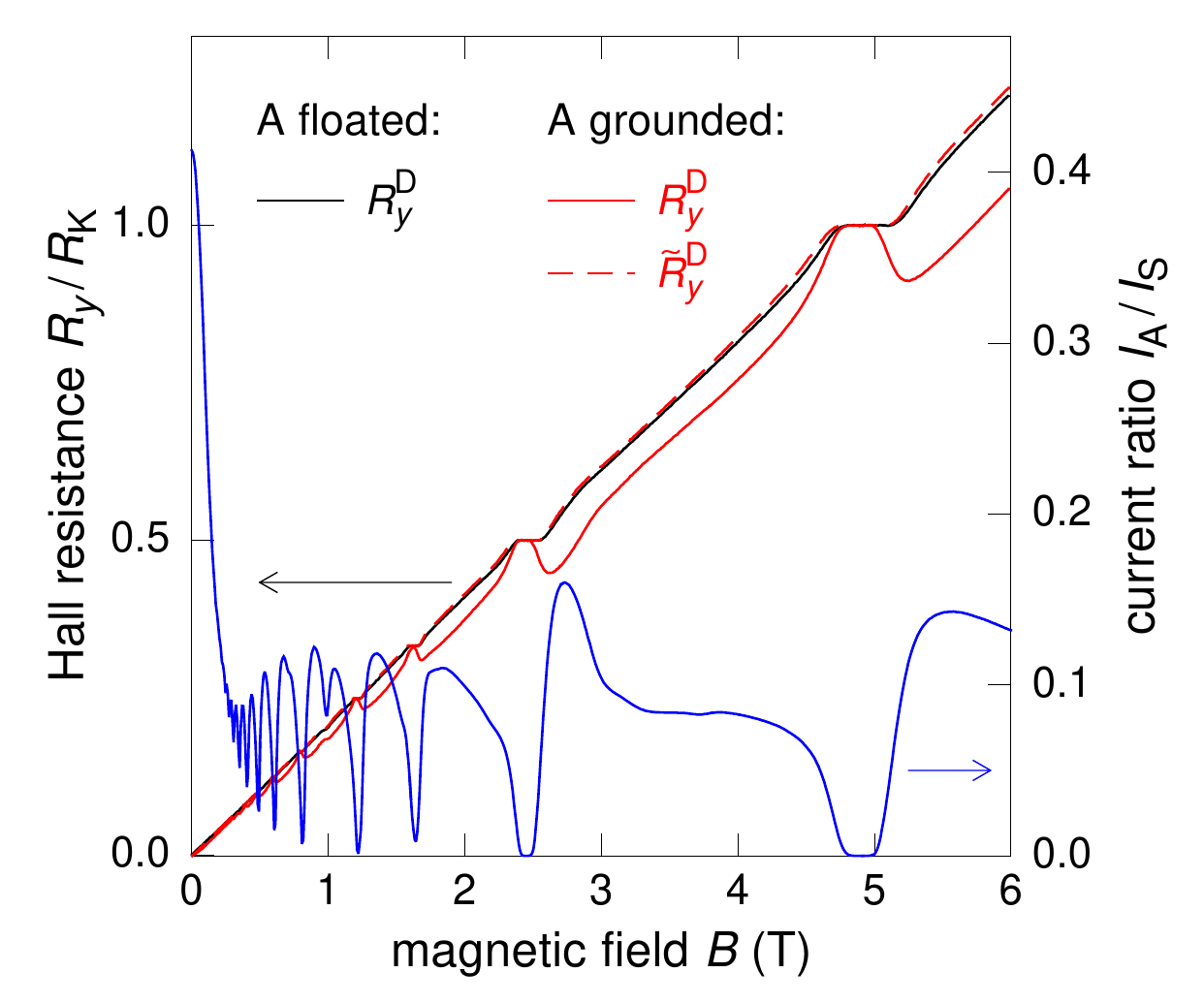}}
\caption{Left axis: measured Hall resistance $R^\mathrm{D}_{y}(B)$ while the inner contact is either floated (black solid line) or grounded (red solid line); the dashed red line is $\widetilde{R}^\mathrm{D}_{y}(B)$, corrected for the reduced current. Right axis: current ratio $I_\mathrm{A}/I_\mathrm{S}$ while A is grounded (solid blue line).}
\label{fig:Agrounded}
\end{figure}
\begin{figure*}[t]
\hspace{-3ex}
\includegraphics[width=.48\textwidth]{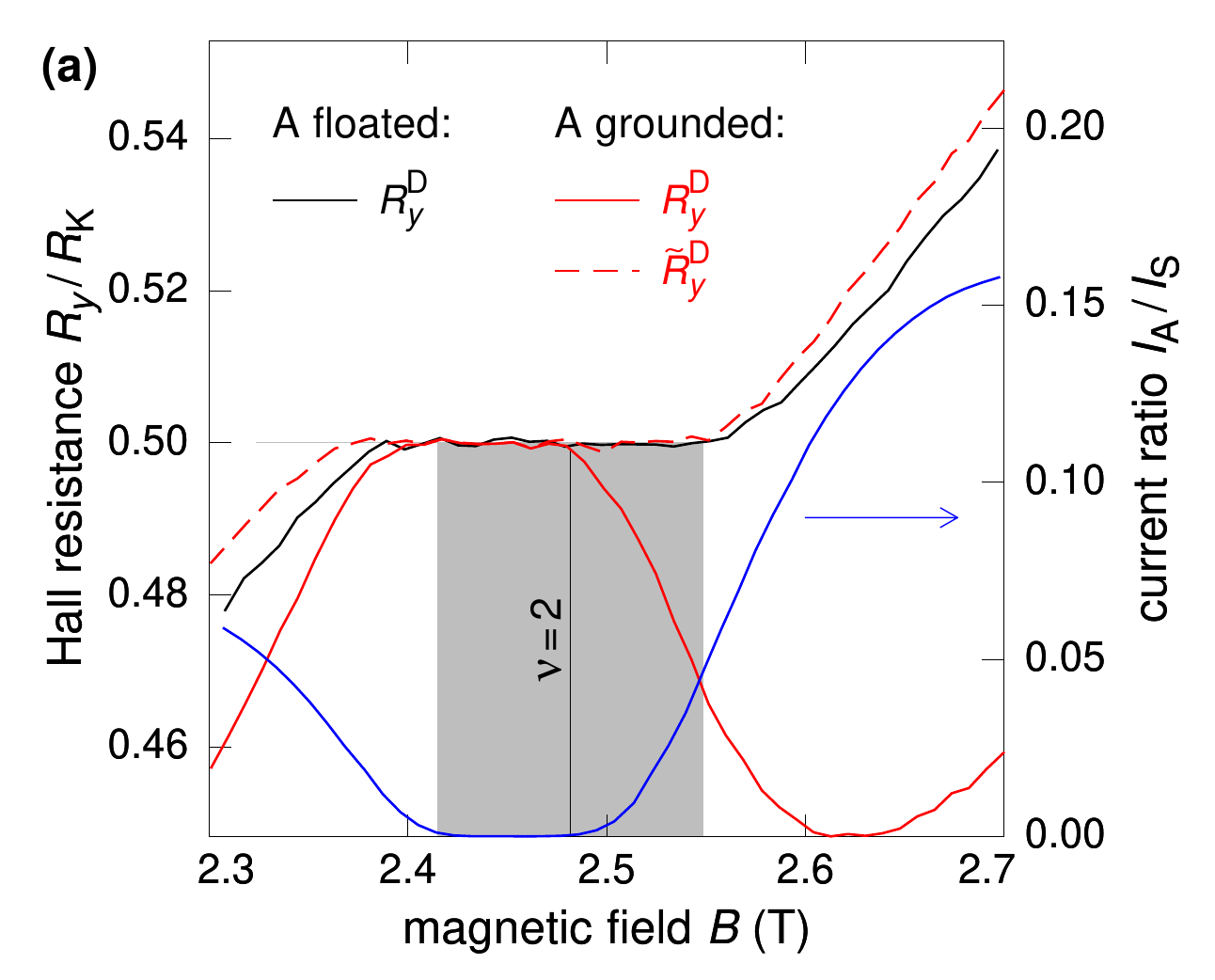}
\includegraphics[width=.48\textwidth]{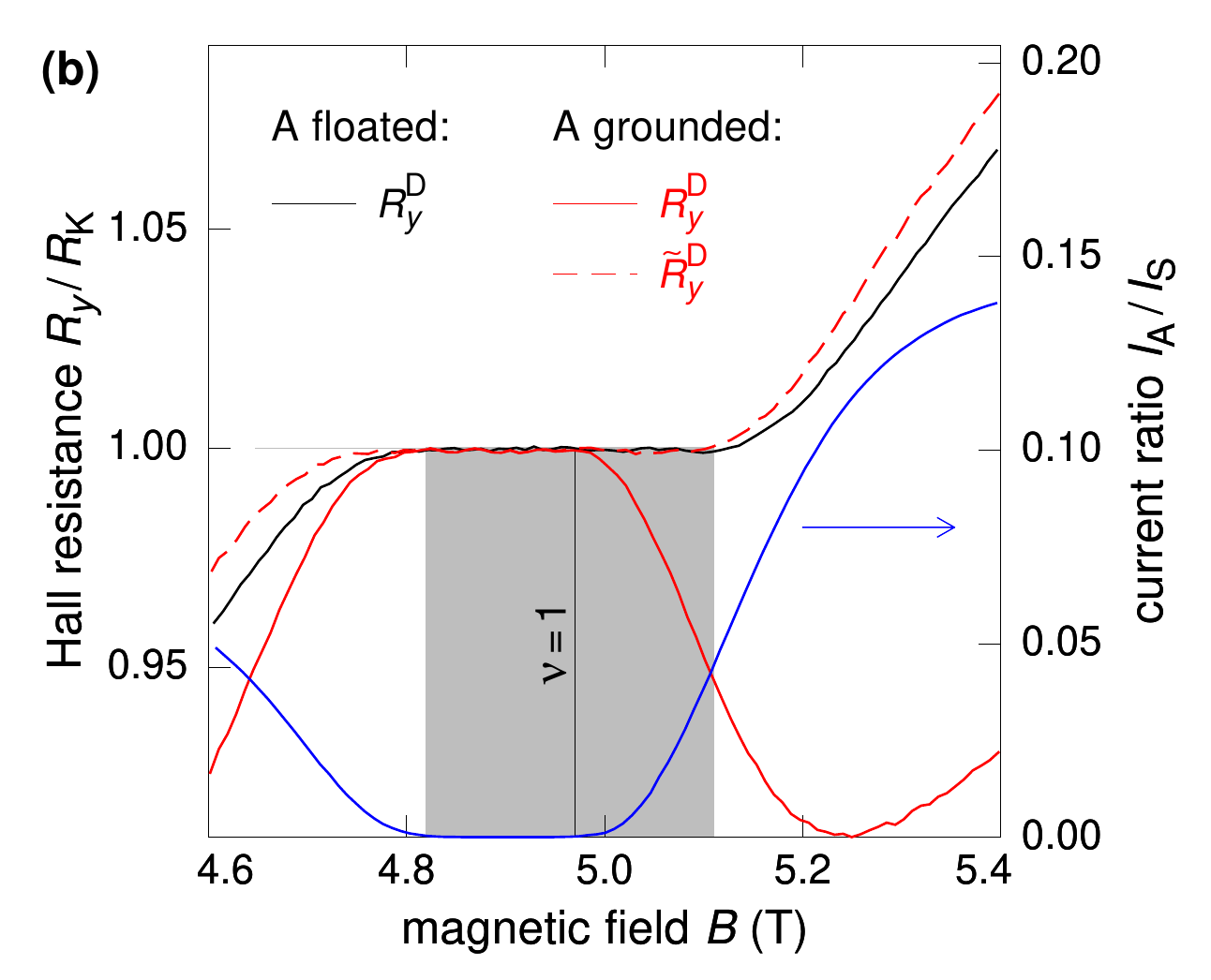}
\caption{Enlargements from Fig.\ \ref{fig:Agrounded} near bulk filling factors $\nu=2$ in (a) and $\nu=1$ in (b), indicated as vertical lines. Magnetic field regions of quantized plateaus are depicted as gray background. On the high-field side of the integer $\nu=1,2$, {$I_\mathrm{A}>0$} while the corrected $\widetilde{R}^\mathrm{D}_{y}$ recovers the quantized plateau resistance.}
\label{fig:Acurrentff12}
\end{figure*}

Our $10\,\upmu\mathrm{m}$ wide Hall bar, presented in Fig.~\ref{fig:sample}(a), is carved from a GaAs/AlGaAs heterostructure containing a two-dimensional electron system (2DES) 130\,nm below the surface. It contains two square shaped inner contacts of size $2.5\,\upmu\text{m}\times2.5\,\upmu\text{m}$ placed in the center of the Hall bar. The inner contacts are electrically connected via air bridges [appearing red in Fig.~\ref{fig:sample}(a)],
such that the Hall bar beneath remains undisturbed \cite{Ji2003}. We cool the sample to a temperature of $T\simeq300\,$mK using a He-3 evaporation cryostat. At this cryogenic temperature, the carrier density and mobility of the 2DES are $n_\mathrm{s} \simeq 1.2\times 10^{11} \mathrm{ cm}^{-2} $ and $ \mu \simeq 4\times 10^6 \mathrm{cm}^2/\mathrm{Vs}$, respectively, as determined from our Hall measurements. The corresponding mean-free-path is $\lambda_\mathrm{m}\simeq23\,\upmu$m.

We impose a constant current of $I_\mathrm{S}=-100\,$nA (using a Keithley 2450 sourcemeter), such that electrons flow through the source contact (S) into the Hall bar ($V_{\mathrm{S}}<0$). The drain contact (D) is connected to the measurement ground ($V_{\mathrm{D}}=0$). All ohmic contacts are equipped with low-pass RC-filters ($R = 2200\,\Omega$, $C = 2\,$nF) for noise reduction. Additional cable resistances ($\sim 100\,\Omega$) and capacitances ($\sim 100$ pF) are much smaller, the resistances of the peripheral ohmic contacts are $\simeq50\,\Omega$. The inner contacts have a higher resistance of $\simeq 1\,\mathrm{k}\Omega$ due to their small size, the left inner contact is always kept floated. This adds up to overall $B$-independent contact resistances of $R_i\simeq2.4\,\mathrm{k}\Omega$ for the peripheral contacts ($i=1,2,3,4,\mathrm{S},\mathrm{D}$) and $R_\mathrm{A}\simeq3.4\,\mathrm{k}\Omega$ for the inner contact, labeled A. At $B=0$, the resistance of the Hall bar between contacts S and D is $R_0\simeq50\,\Omega$.

Contacts 1--\,4 serve as voltage probes. We simultaneously measure the four individual voltages $V_{1,2,3,4}$ in respect to ground (using Agilent 34411A multimeters). \footnote{Compared to measuring pairwise voltage differences, this increases flexibility but slightly reduces accuracy. To explore the highest possible accuracy of our multimeters we calibrated them using the measured Hall resistances of the quantized plateaus. The corrections are $<1\%$.}

Defining $R_{ij}=(V_j - V_i)/I_\mathrm{S}$, we label the Hall resistances as $R^\mathrm{S}_{y}\equiv R_{14}$ near the source contact and $R^\mathrm{D}_{y}\equiv R_{23}$ near drain. Likewise, the longitudinal resistances are  $R^\mathrm{high}_{x}\equiv R_{12}$ and $R^\mathrm{low}_{x}\equiv R_{43}$ measured along the high- versus low-potential edges of the Hall bar. In regard to the electron flow $R^\mathrm{S}_{y}$ is measured upstream and $R^\mathrm{D}_{y}$ downstream of the inner contacts.

If we keep both inner contacts electrically floated, such that electrons flow from S to D, we find $R^\mathrm{S}_{y}=R^\mathrm{D}_{y}$ and $R^\mathrm{low}_{x}=R^\mathrm{high}_{x}$. In particular, this result confirms that the inner contacts have a negligible effect on the Hall measurements in a high magnetic field, if they are floated.

If we connect A to ground, part of the current, labeled $I_\mathrm{A}$, flows through it. In Fig.~\ref{fig:Agrounded}{}, we show the Hall resistance $R^\mathrm{D}_{y}(B)$ near the drain contact (left axis) for A floated (black line) versus A grounded (red solid line). Grounding A yields a reduction of $R^\mathrm{D}_{y}$ whenever $I_\mathrm{A}\ne0$; the blue line in Fig.~\ref{fig:Agrounded}{} displays the current ratio $I_\mathrm{A}(B)/I_\mathrm{S}$ (right axis). We observe a sharp drop of $I_\mathrm{A}(B)/I_\mathrm{S}$ for $B\lesssim0.3\,$T, i.e., in the classical regime with $R_y(B)\propto B$.
An even steeper drop of $I_\mathrm{A}(B)$ is expected for a two-terminal Corbino geometry, consisting of two contacts separated by a conducting plane without sample edges between the contacts. Our three-terminal device has an inner contact surrounded by 2DES, too. However, unlike a Corbino geometry, it contains two outer contacts at different chemical potentials ($\mu_\mathrm{S}$ versus $\mu_\mathrm{D}$). In our experiments, for $B>0.3$\,T the main share of the current flows between the outer contacts. The sample edges connecting them charge up accordingly and yield an in-plane electric Hall field perpendicular to the current density.

\begin{figure*}[th]
\hspace{-40mm}{\includegraphics[width=0.55\textwidth]{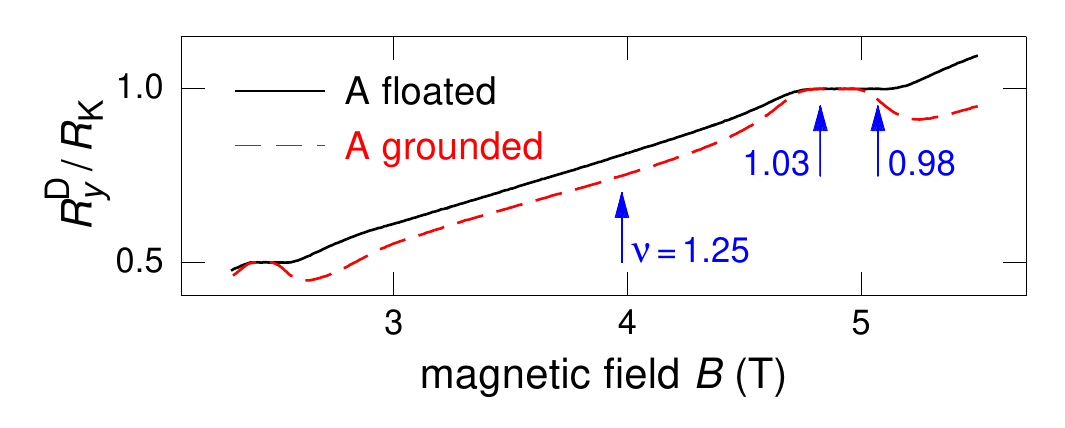}}\\%
\centerline{\includegraphics[width=.8\textwidth]{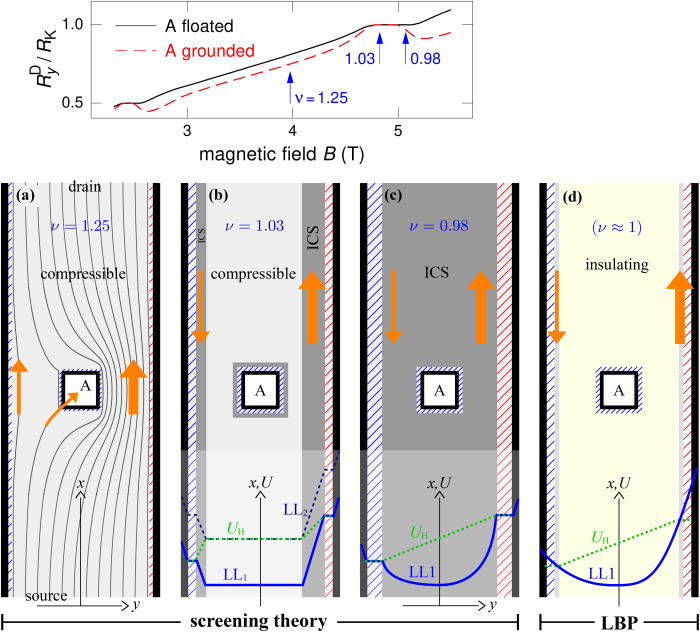}}%
\caption{\textbf{Screening theory (a-c):}
Qualitative sketches of the segmentation of the Hall bar in compressible (light gray) and incompressible (dark gray) regions for contact A connected to ground at three different filling factors near $\nu=1$ as indicated in the top panel for orientation. The insulating sample edges and a thin Schottky (tunnel) barrier around A are  drawn as black lines. The low (high) potential edges are indicated by blue (red) stripe patterns. Arrows (orange) indicate the local flow of electrons.
(a) Diffusive regime between plateaus; potential drops gradually across the compressible Hall bar; the current flows uni-directionally along the (thin black) equipotential lines.
(b, c) Plateau region; chiral current flow restricted to ICSs; potential changes linearly inside ICSs and is constant in compressible regions; (b) features the case of egde ICSs, which in (c) widened and combined to a single bulk ICS.
\textbf{LBP (d):} Situation assumed in the LBP for the entire plateau. Current is restricted to edge channels (light gray), insulating bulk (yellow).
Line plots at bottom: LL energies across the Hall bar (blue lines solid for LL$_1$ and dashed for LL$_2$) and the corresponding Hall potential $U_\mathrm{H}(y)$ (dotted green lines). LLs are completely filled at energies below $U_\mathrm{H}(y)$, partly filled at $U_\mathrm{H}(y)$ and empty above $U_\mathrm{H}(y)$.
}
\label{fig:sketch}
\end{figure*}

In Ref.\ \textbf{[supplementary material]} we present a complete set of measurements and offer a detailed explanation of the current flow into the inner contact. In short, contrary to a Corbino device, the chemical potential of the inner contact fulfills $\mu_\mathrm{D}<\mu_\mathrm{A}\le0.5\left(\mu_\mathrm{S}+\mu_\mathrm{D}\right)$. Therefore, we expect $I_\mathrm{A}>0$ even for $\mu B\gg1$, where the current flows along the equipotential lines. [We visualize the latter in Fig.\ \ref{fig:sketch}(a), below.]

For $B>0.3\,$T, the current ratio $I_\mathrm{A}(B)/I_\mathrm{S}$ features pronounced Shubnikov-de-Haas oscillations [related to the oscillation of $\mu(B)$] averaging to $\overline{I_\mathrm{A}}\simeq0.08I_\mathrm{S}$. The quantized plateaus of the Hall resistance indicate $\mu B\to\infty$, beyond the scope of the Drude model. For the largest part of the plateaus, we observe $I_\mathrm{A}=0$, indicating a complete electrical isolation of the inner contact.

In the case of $I_\mathrm{A}\ne0$, we account for the reduction of the current to $I_\mathrm{S}\to I_\mathrm{S}-I_\mathrm{A}$ downstream of A by introducing a corrected Hall resistance $\widetilde{R}^\mathrm{D}_{y}=R^\mathrm{D}_{y}\,\frac{I_\mathrm{S}}{I_\mathrm{S}-I_\mathrm{A}}$, which we added in Fig.~\ref{fig:Agrounded}{} as a red dashed line. At large, $\widetilde{R}^\mathrm{D}_{y}(B)$ recovers $R^\mathrm{D}_{y}(B)$ measured with A floated. However, a detailed look reveals distinct deviations suggesting an interesting dynamics.

Focusing on the quantum Hall regime, in Fig.~\ref{fig:Acurrentff12}{} we display enlarged sections of Fig.~\ref{fig:Agrounded}{} covering the first two plateaus near bulk filling factors $\nu=1$ and $\nu=2$ (averaged over the width of the Hall bar). The shaded backgrounds indicate the plateau regions measured for floated A (with $R^\mathrm{D}_{y}=\frac1\nu h/e^2$ and $R^\mathrm{low}_{x}=R^\mathrm{high}_{x}=0$). The observed behavior is congruent for $\nu=1,2,\dots$.

Throughout the low magnetic field side of the plateaus, the grounded inner contact is electrically isolated with $I_\mathrm{A}=0$ and leaves the quantized plateaus unaffected, i.e., $R^\mathrm{D}_{y}=R_\mathrm{K}/\nu$. However, if we increase {$B$} while remaining on the quantized plateau for $\nu\le1$ or $\nu\le2$, a rapidly growing fraction of the current $I_\mathrm{S}$ flows through the inner contact and $R^\mathrm{D}_{y}<R_\mathrm{K}/\nu$. Nevertheless, the corrected $\widetilde{R}^\mathrm{D}_{y}$ exactly recovers the quantized plateau resistance.

\begin{figure}[t]
\centerline{\includegraphics[width=1\columnwidth]{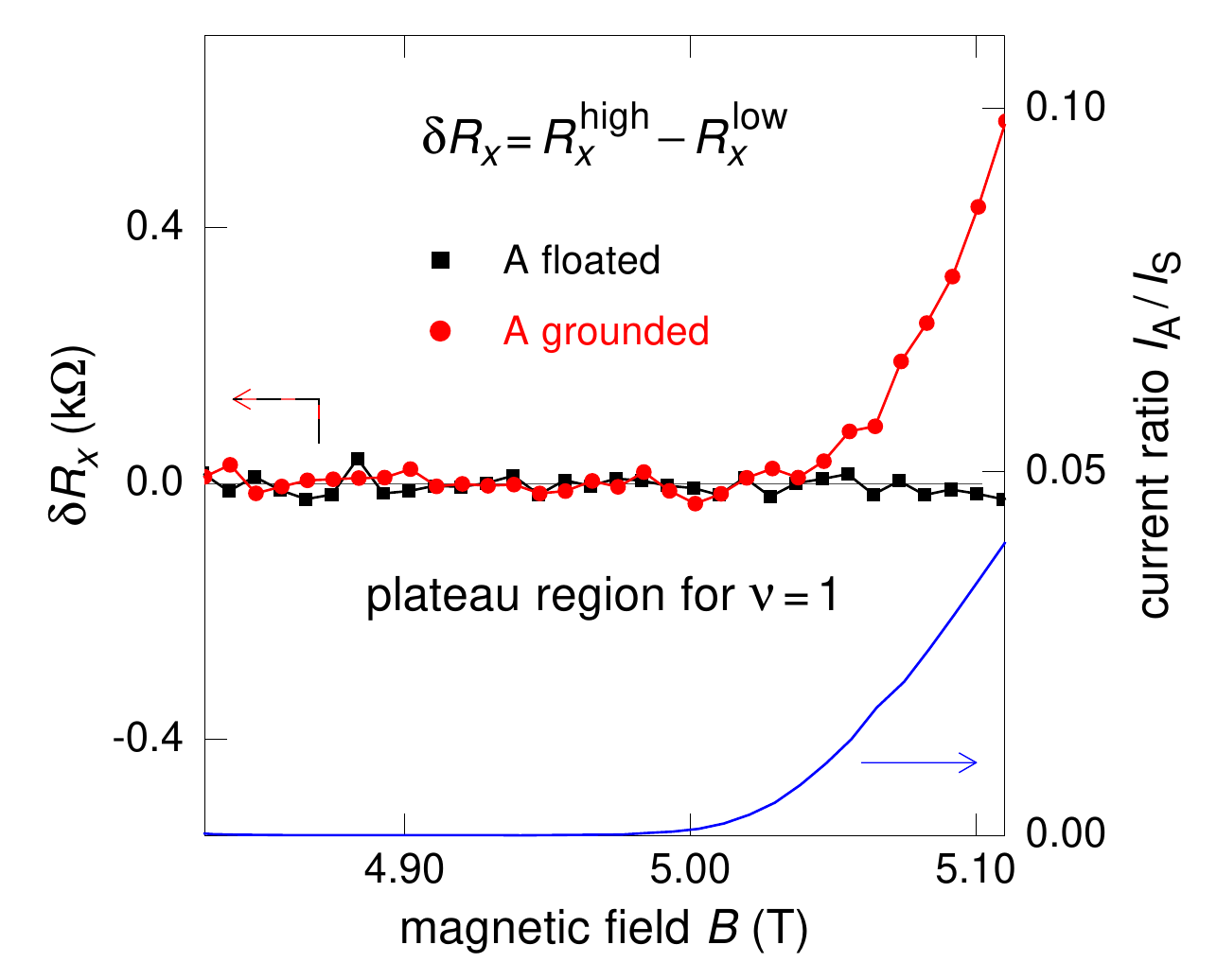}}
\caption{Left $y$-axis: difference $\delta R_x$ between longitudinal resistance on high versus low potential sides of the Hall bar in the region of the first plateau ($\nu=1$) for contact A grounded (solid red line) versus A floated (dashed black line). The horizontal line indicates $\delta R_x=0$.
Right $y$-axis: current ratio $I_\mathrm{A}/I_\mathrm{S}$ for A grounded.}
\label{fig:volt_diff}
\end{figure}
In Fig.~\ref{fig:volt_diff}, we confirm that current flowing into the inner contact also yields a drop of the chemical potential along the high potential side of the Hall bar, i.e., a finite $I_\mathrm{A}$ yields a longitudinal resistance $R^\mathrm{high}_{x}>0$. We compare the onset of $I_\mathrm{A}$ along the  $\nu=1$ plateau with the difference $\delta R_x=R^\mathrm{high}_{x}-R^\mathrm{low}_{x}$. For floated A, $I_\mathrm{A}=\delta R_x=0$. However, for grounded A, $\delta R_x>0$ for $I_\mathrm{A}>0$. We find $R^\mathrm{high}_{x}-R^\mathrm{low}_{x}=R^\mathrm{S}_{y}-R^\mathrm{D}_{y}$ in accordance with Kirchhoff's voltage law.


%
Our experiments clearly demonstrate that, starting in the middle of the plateau, current flows in the bulk of the Hall bar. Even for $I_\mathrm{A}\ne0$, the QHE does not break down, instead we find identical plateau regions at quantized Hall resistance $R_\mathrm{K}/\nu$, whether the inner contact is grounded or floated, cf.\ Fig.~\ref{fig:Agrounded}{}. We interpret this finding in terms of scattering-free bulk-current as follows: Whenever $I_\mathrm{S}$ reaches the inner contact, the drop of the chemical potential across the current carrying ICS is reduced by $e|I_\mathrm{A}|R_\mathrm{K}/\nu$ according to $I_\mathrm{A}$ branching off. This transition is adiabatic, meaning that electron scattering inside the ICS remains completely suppressed while the potential distribution changes smoothly.

Our observations are in contradiction to the LB picture in two ways: First, within the LB picture current flow into the inner contact would be associated with a breakdown of the QHE, which we do not observe, cf.\ Sec.\ IV of Ref.\ \textbf{[supplementary material]}. (Instead, we find that $\widetilde{R}^\mathrm{D}_{y}$ exactly recovers the quantized plateau resistance.) Second, the LB picture does not provide a scenario, which breaks the symmetry of the plateaus. This is incompatible with our observation that $I_\mathrm{A}\ne0$ for the higher-magnetic-field halfs of the plateaus while $I_\mathrm{A}=0$ otherwise.

We observe a gradual transition from edge-currents to scattering-free bulk-current as the magnetic field is increased along a quantized plateau. This transition causes the observed asymmetry of $I_\mathrm{A}(B)$ and $R^\mathrm{D}_{y}(B)$ in respect to the integer bulk filling factor. Our findings point to the importance of the Coulomb interaction between the charge carriers as predicted by the screening theory \cite{Chklovskii1992,Gerhardts2008}.  In Fig.~\ref{fig:sketch}{}, we qualitatively sketch the prediction of the screening theory for three different filling factors. In Fig.~\ref{fig:sketch}(a), we show the situation for $\nu\simeq1.25$ depicting the diffusive transport regime between plateaus (Drude model), in Fig.\ \ref{fig:sketch}(b) and \ref{fig:sketch}(c), we sketch typical situations along the quantized plateau, cf.\ Ref.\ \onlinecite{Yildiz2014} for comparable numerical calculations. In Fig.~\ref{fig:sketch}(b) with $\nu\simeq1.03$, we consider the low-magnetic-field side of the plateau, where separate narrow ICSs follow along the sample edges. In Fig.~\ref{fig:sketch}(c) with $\nu\simeq0.98$ at the high-field side of the plateau, a single ICS extends through the bulk of the Hall bar. Compressible regions are indicated with light gray shading, ICSs with dark gray shading. Colored striped patterns mark the compressible regions at the highest (red) and lowest (blue) potentials.

In the diffusive regime in Fig.~\ref{fig:sketch}(a), the entire Hall bar is compressible, the Hall potential $U_\mathrm{H}$ drops across its entire width and current flows everywhere. Equipotential lines are indicated by thin lines current flow by orange arrows. In the case of a quantized plateau as shown in Figs.\ \ref{fig:sketch}(b) and \ref{fig:sketch}(c), $U_\mathrm{H}$ drops entirely inside the ICSs, which are indicated using dark gray shading. We plot these Hall potentials $U_\mathrm{H}(y)$ together with the course of the $\nu=1$ LL energy in the bottom regions of Figs.\ \ref{fig:sketch}(b) and \ref{fig:sketch}(c). The current flows where the potential drops, hence, for the plateaus it is restricted to the ICSs. For the case of edge ICSs, cf.\ Fig.~\ref{fig:sketch}(b), the inner contact is isolated and $I_\mathrm{A}=0$. However, in the diffusive regime, cf.\ Fig.~\ref{fig:sketch}(a), or if a bulk ICS exists, cf.\ Fig.~\ref{fig:sketch}(c), current can flow into contact A.

The screening theory predicts that an increase of $I_\mathrm{S}$ results in a widening of the ICS at the high potential edge of the Hall bar (which then carries more current), cf.\ Fig.\ \ref{fig:sketch}(b). We expect that a wider ICS (for a larger $I_\mathrm{S}$) reaches the inner contact and yields $I_\mathrm{A}\ne0$ already at smaller $B$. We present measurements confirming this behavior in Sec.\ IV of the Ref.\ \textbf{[supplementary material]}.

In the LB picture, the bulk of the Hall bar is assumed to be insulating, such that current can flow only inside the compressible 1D edge channels, cf.\ Fig.~\ref{fig:sketch}(d). Our observation of $I_\mathrm{A}\ne0$ for quantized plateaus clearly support the screening theory but contradict the LB picture.


In summary, our Hall bar equipped with an inner contact allows us to directly measure whether current flows in the bulk of the sample. Our results indicate an asymmetry along the quantized plateaus of the Hall resistance: For the low-magnetic-field side of the plateaus the current through the inner contact is $I_\mathrm{A}=0$, suggesting that the applied current is restricted to the edges of the Hall bar. In contrast, for the high-magnetic-field side of the plateau ($\nu$ lesser than the integer value), we observe $I_\mathrm{A}\ne0$, where our analysis indicates scattering-free bulk-current. These findings are incompatible with the LB edge channel picture. Instead, they confirm the relevance of electron-electron interactions for the QHE. Our results can be qualitatively explained within the screening theory, which includes the direct Coulomb interaction between the carriers. A quantitative comparison would require extended numerical calculations, which is beyond the scope of this letter. 

Our findings indicate the possible importance of the direct Coulomb interaction for other types of the QHE, including its topological variants. Moreover, the more complicated current density distribution predicted by the screening theory (compared to the LB picture) should be considered for experiments in the QHE regime, particularily for the case of phase sensitive measurements. Finally, the application of the QHE as resistant standard might be further optimized, if we manage to apply our improved microscopic understanding to create Hall bars with more stable plateaus.

\section*{Supplementary Material}
The supplementary material contains a complete set of Hall measurements with the inner contact floated or grounded, respectively, as well as a detailed explanation of the current flow into the inner contact including a comparison with the Corbino geometry. We also present additional data exploring the breakdown of the quantum Hall effect. They indicate that the inner contact current presented in the main article is not related with the breakdown of the quantum Hall effect.

\section*{Acknowledgements}

This work was funded by the Deutsche Forschungsgemeinschaft (DFG, German Research Foundation) -- 218453298. The authors thank Lutz Schrottke for helpful discussions.

\section*{Author contributions}

The authors declare no competing interests. V.\,U.\ produced the sample. S.\,S.\ performed the measurements. S.\,S.\ and S.\,L.\ analyzed the data and wrote the article. A.\,S.\ contributed to the planning and provided critical feedback. S.\,L.\ led the project.

\section*{Data Availability}

The data that support the findings of this study are available from the corresponding author upon reasonable request.

\section*{References}
%

\end{document}


\title{Supplementary Material:\\ Transition from edge- to bulk-currents in the quantum Hall regime}

\author{S. Sirt}\email{sirt@pdi-berlin.de}\affiliation{Paul-Drude-Institut f\"ur Festk\"orperelektronik, Leibniz-Institut im Forschungsverbund Berlin e.V., Hausvogteiplatz 5-7, 10117 Berlin, Germany}
\author{V.\,Y. Umansky}\affiliation{Weizmann Institute of Science, 234 Herzl Street, POB 26, Rehovot 76100, Israel}
\author{A. Siddiki}\affiliation{Istanbul Atlas University, Hamidiye, Anadolu Cd. no:40, 34408, Istanbul, Turkey}
\author{S. Ludwig}\email{ludwig@pdi-berlin.de}\affiliation{Paul-Drude-Institut f\"ur Festk\"orperelektronik, Leibniz-Institut im Forschungsverbund Berlin e.V., Hausvogteiplatz 5-7, 10117 Berlin, Germany}

\maketitle

\renewcommand{\thefigure}{S\arabic{figure}}

\section{Complete set of Hall measurements}\label{app:resistances}

\begin{figure}[th]
\includegraphics[width=1\columnwidth]{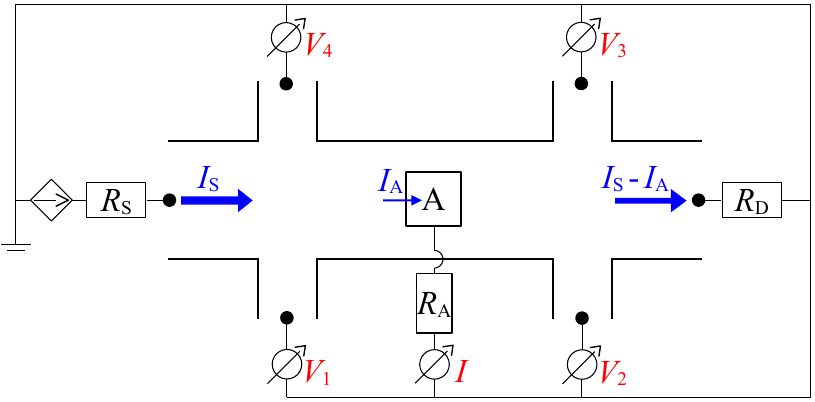}%
\caption{Sketch of the sample with a simplified circuit diagram of the set-up. The current $I_\mathrm{S}$ is applied (and measured) using a constant current source. $I_\mathrm{A}$ flowing via the contact resistance $R_\mathrm{A}$ to ground is measured as well, if contact A is grounded. The reminder of the current flows via the contact resistance $R_\mathrm{D}$ to ground. The voltage probes 1--4 are kept electrically floating, while the voltages $V_{1,2,3,4}$ are simultaneously measured with four separate multimeters.}
\label{fig:sketch_suppl}
\end{figure}

\begin{figure*}[th]
\includegraphics[width=1\columnwidth]{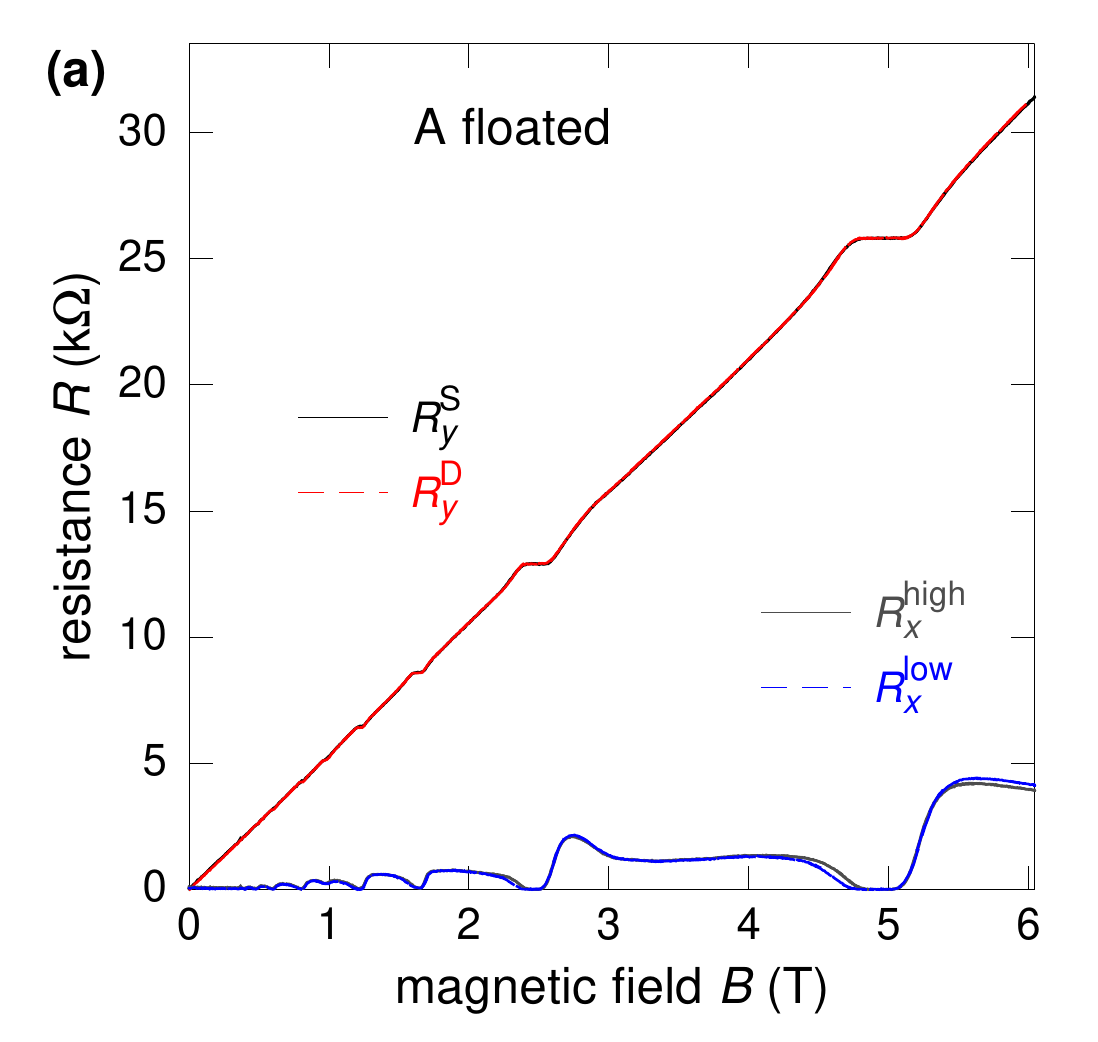}%
\hspace{2ex}
\includegraphics[width=1\columnwidth]{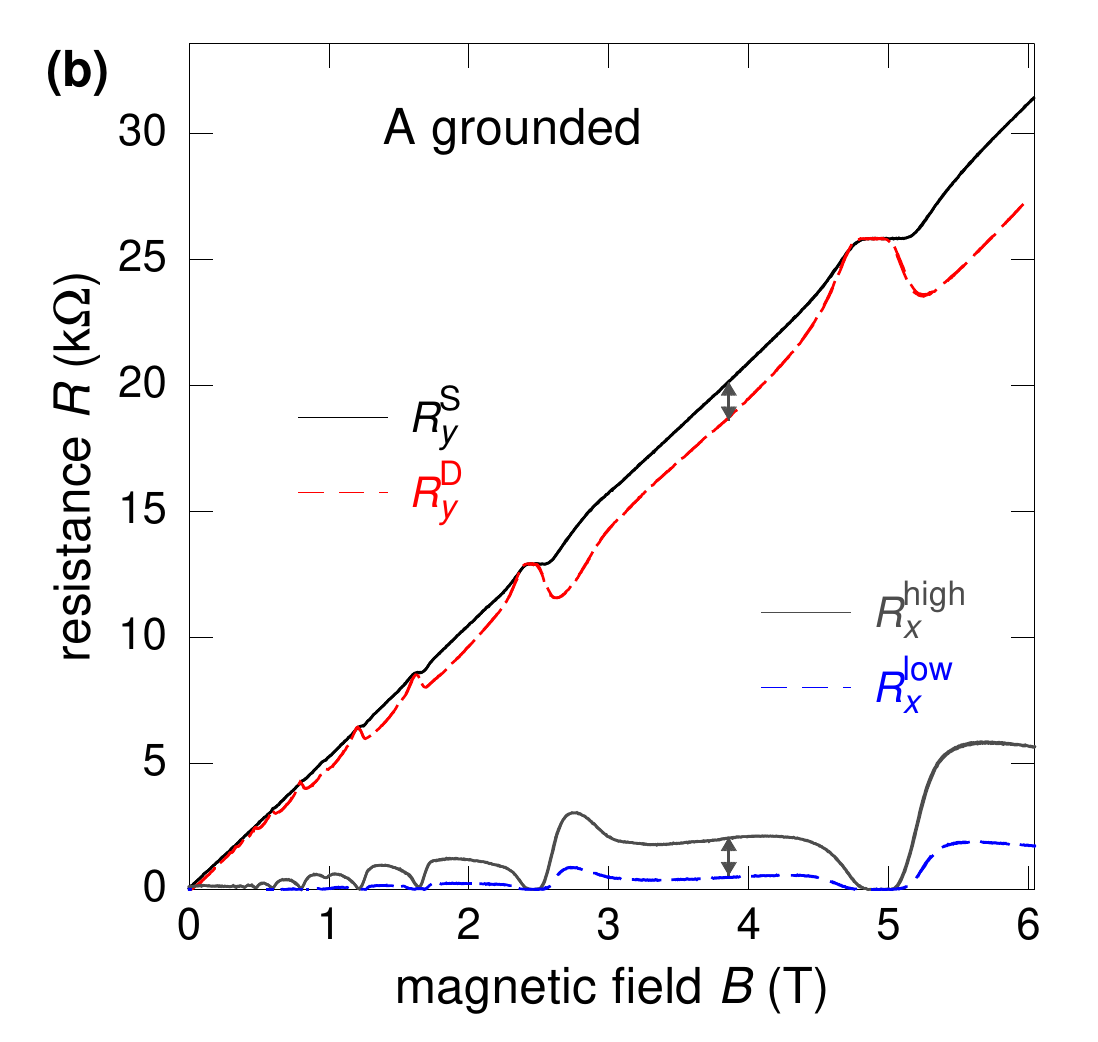}%
\caption{Hall resistances $R_y$ and longitudinal resistances $R_x$ while the inner contact A is kept electrically floated (a) or connected to ground (b). The two gray double arrows in (b) have identical length.}
\label{fig:all_resistances}
\end{figure*}

\begin{figure}[th]
\includegraphics[width=.96\columnwidth]{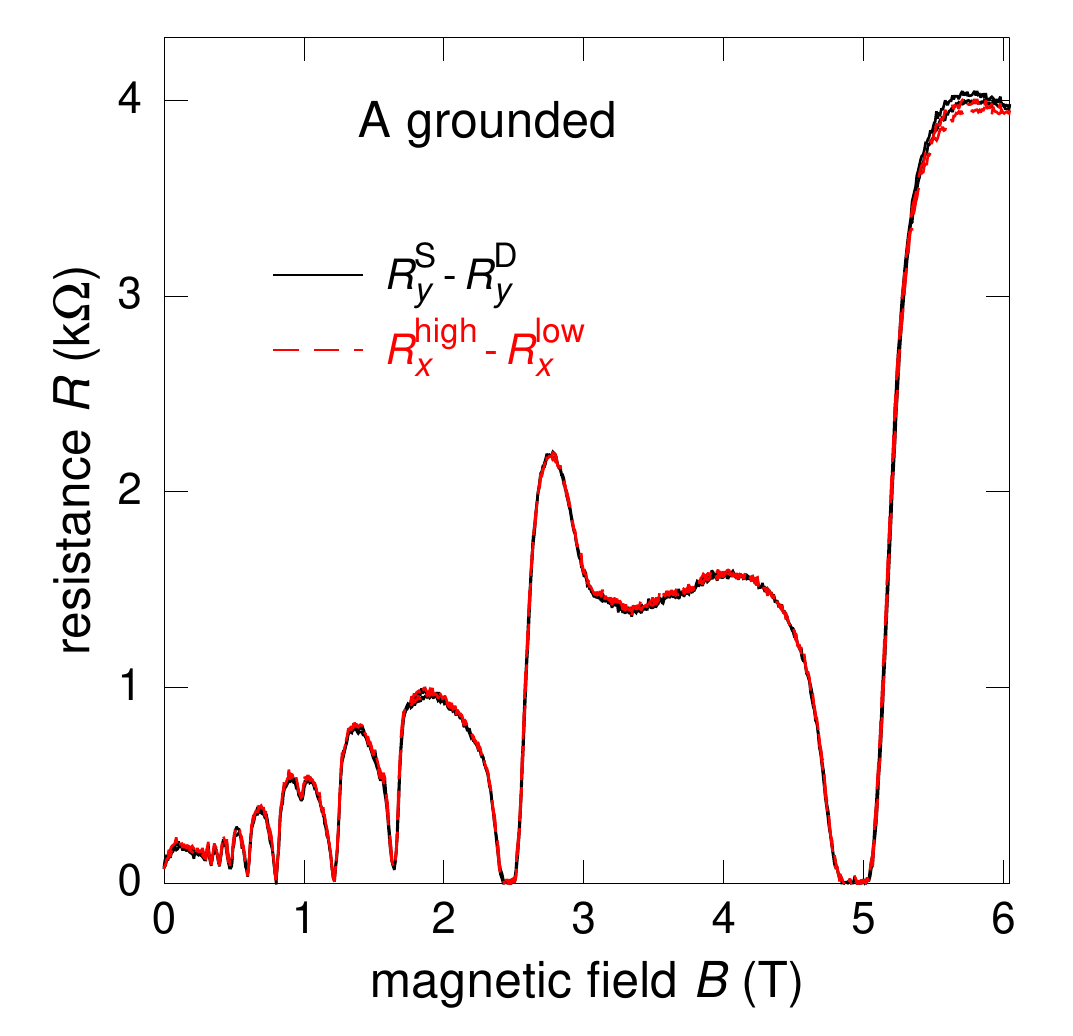}%
\caption{The inner contact A is connected to ground. The difference between the Hall resistances matches the difference between the longitudinal resistances. After multiplying the resistances with $I_\mathrm{S}$, this confirms Kirchhoff's voltage law stating that the direct sum of potential differences around a closed loop is zero.}
\label{fig:A-subtracted_app}
\end{figure}

In Fig.~\ref{fig:sketch_suppl}, we sketched the sample including a simplified circuit diagram of our measurement set-up.
In Fig.~\ref{fig:all_resistances}, 
we plot both, Hall and longitudinal resistances as a function of $B$, while the inner contact A is electrically floated [panel (a)] in comparison to the case when A is connected to ground [panel (b)]. While A is electrically floated, we find the expected relations $R^\mathrm{S}_{y}=R^\mathrm{D}_{y}$ and $R^\mathrm{high}_{x}=R^\mathrm{low}_{x}$, suggesting that the floated inner contacts do not alter the Hall effect. \footnote{Tiny deviations in the measurement (beyond statistical noise) are caused by a slightly imperfect calibration of the four separate voltmeters.} If we ground the inner contact, $R^\mathrm{S}_{y}$ remains unchanged but we observe $R^\mathrm{D}_{y}\le R^\mathrm{S}_{y}$ and $R^\mathrm{low}_{x}\le R^\mathrm{high}_{x}$. The potential differences of a closed loop sum up to zero corresponding to $R^\mathrm{S}_{y}-R^\mathrm{D}_{y}=R^\mathrm{high}_{x}-R^\mathrm{low}_{x}$ for our Hall bar. (Remember, that we defined all these ``resistances'' by dividing the measured voltage drops by the identical current $I_\mathrm{S}$.) This is experimentally confirmed in Fig.~\ref{fig:A-subtracted_app}, suggesting that the four individual voltmeters are correctly calibrated. \footnote{We compared the quantized Hall resistances at the plateaus with the von-Klitzing constant in order to control and, if necessary, fine-tune the calibration of the voltmeters.}

If the inner contact is grounded, we find in the diffusive limit, i.e., between plateaus for the longitudinal voltage drops $R^\mathrm{low}_{x}<R^\mathrm{high}_{x}$. In fact, compared to the case of A being floated in Fig.~\ref{fig:all_resistances}(a), for A being grounded in Fig.~\ref{fig:all_resistances}(b), the current density is reduced on the low-potential side of the Hall bar while it is increased on its high-potential side. The latter can be understood in terms of the reduced Hall voltage downstream of A due to $I_\mathrm{S}\to I_\mathrm{S}-I_\mathrm{A}$, which effects the high potential edge of the Hall bar. In comparison, $R^\mathrm{low}_{x}$ along the low potential edge of the Hall bar should not be affected by the reduced Hall voltage but should only be slightly reduced (in the order of $10\%$) due to $I_\mathrm{S}\to I_\mathrm{S}-I_\mathrm{A}$. Instead, $R^\mathrm{low}_{x}$ is substantially reduced. It points towards an inhomogeneous and asymmetric current distribution around the grounded inner contact. We study the effect of the inner contact more closely in Sec.~\ref{app:model} below.

\section{Current flowing through the inner contact}\label{app:inner_current}

\begin{figure}[th]
\includegraphics[width=1\columnwidth]{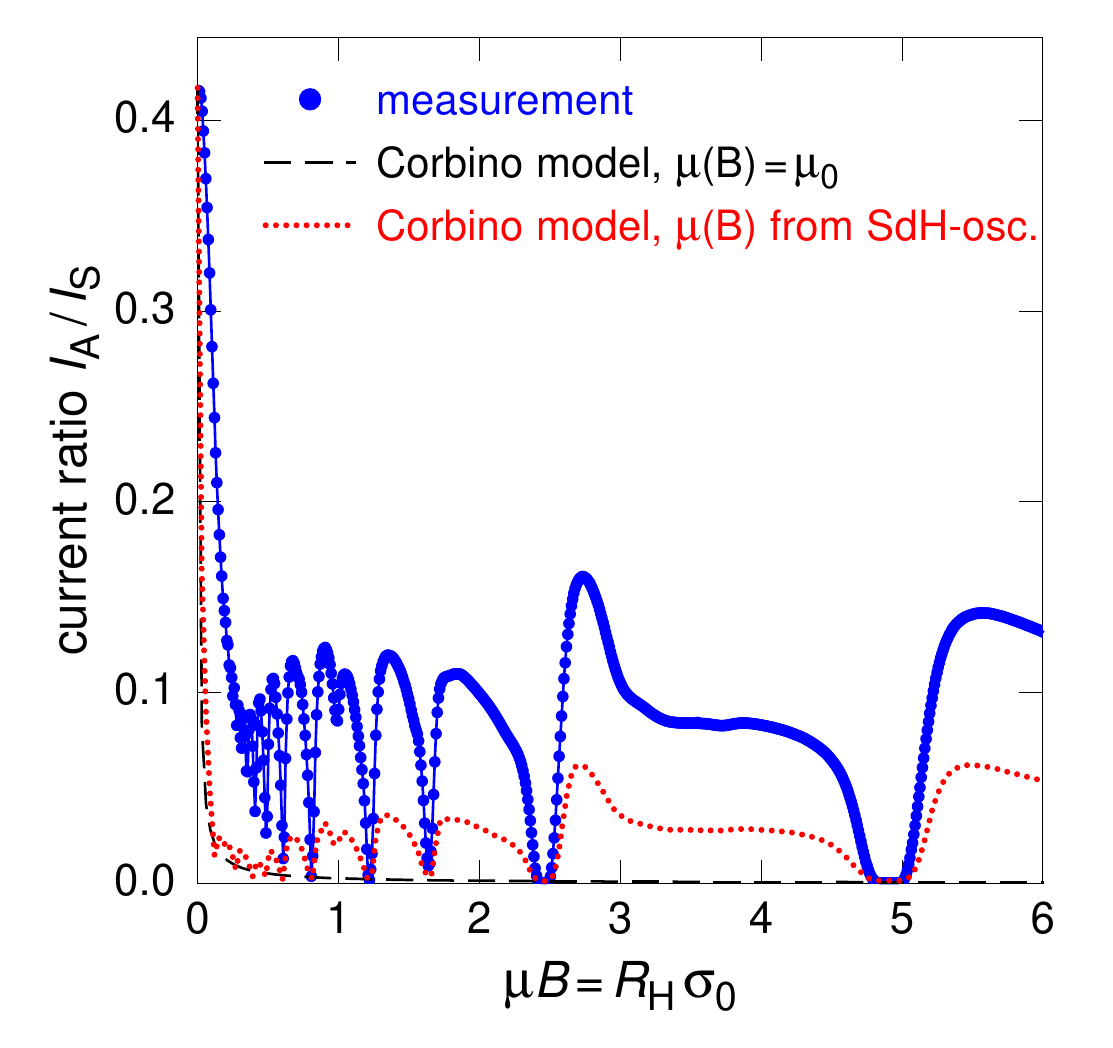}%
\caption{Ratio of the currents flowing through the inner contact and the source contact $I_\mathrm{A}/I_\mathrm{S}$ as a function of $B$. The dashed black and red dotted lines are model curves assuming a Corbino geometry, cf.\ main text, for a constant mobility $\mu_0=\mu(B=0)=396\,\mathrm{m}^2/\mathrm{Vs}$ or the strongly oscillating $\mu(B)$ determined from the longitudinal resistance measurements with the inner contact floated, respectively.}
\label{fig:inner_current}
\end{figure}

In Fig.~\ref{fig:inner_current},
we present the measured current $I_\mathrm{A}(\mu B)$ flowing through the inner contact, which is also shown in Fig.\ 2 of the main article. If we consider the enlargements shown in Fig. 3 of the main article in addition, we observe four different principle regimes of $I_\mathrm{A}(B)$.
\renewcommand{\theenumi}{\roman{enumi}}
\begin{enumerate}
 \item \textit{Scattering-free edge current}: within the lower-magnetic-field regions of the plateaus, the entire current flows without scattering through ICSs along the edges of the Hall bar. The inner contact is completely isolated and $I_\mathrm{A}=0$, cf.\ Fig. 3 of the main article.
 \item \textit{Scattering-free bulk current}: within the higher-magnetic-field regions of the plateaus, the current still flows without scattering in an ICS, which now covers the bulk of the Hall bar. A fraction of the current flows through the inner contact, cf.\ Fig. 3 of the main article.
 \item \textit{Classical regime for $B<0.3\,\mathrm{T}$}: at $B=0$, about 42\,\% of $I_\mathrm{S}$ flows through the inner contact related with the ratio of the contact and filter resistances $R_\mathrm{D}\simeq2.4\,\mathrm{k}\Omega$ and $R_\mathrm{A}\simeq3.4\,\mathrm{k}\Omega$. As the magnetic field is increased, $I_\mathrm{A}/I_\mathrm{S}$ drops to about 8\,\% in average.
 \item \textit{Diffusive regime for $B>0.3\,$T}: the Landau quantization causes an oscillation of $I_\mathrm{A}/I_\mathrm{S}$ around the average value of about 8\,\%. This oscillation includes the regions of quantized Hall plateaus, regimes i.\ and ii.\ above. It also includes regions of diffusive current between the Hall plateaus shaping the local maxima of $I_\mathrm{A}$.   
\end{enumerate}

In Fig.~\ref{fig:inner_current} we compare $I_\mathrm{A}(B)/I_\mathrm{S}$ with the classical prediction of the Drude model assuming a Corbino geometry centered around the inner contact, $I_\mathrm{A}(B)/I_\mathrm{S}=\sqrt{(I_\mathrm{A}(0)/I_\mathrm{S})^2+(\mu B)^2}/[1+(\mu B)^2]$. For $\mu B\gg1$ the Hall voltage between the Hall-bar edge and the inner contact dominates $I_\mathrm{A}$, while the current $I_\mathrm{A}(0)/I_\mathrm{S}$ measured at $B=0$ determines the ohmic resistances. The dashed black line is the model prediction, if we assume a constant mobility $\mu_0=\mu(B=0)=396\,\mathrm{m}^2/\mathrm{Vs}$. However, in reality, $\mu(B)$ strongly oscillates related with the Shubnicov-de-Haas oscilllations of the longitudinal resistance. The red dotted line in Fig.~\ref{fig:inner_current} shows our model prediction using the oscillating $\mu(B)$ determined from the measured longitudinal resistance $R_x(B)$, while the inner contact was floated. For all magnetic fields $B>0$ with $I_\mathrm{A}(B)/I_\mathrm{S}\ne0$ our actually measured current ratio $I_\mathrm{A}(B)/I_\mathrm{S}$ substantially exceeds the prediction for a Corbino geometry. Clearly, the two-terminal Corbino model does not correctly describe our three-terminal experiment. In the next section, we explain this deviation.

\section{Explanation of current flow into the inner contact}\label{app:model}

In this section, we present a qualitative model of our three-terminal setup and compare it with the two-terminal Corbino geometry. We restrict our discussion to the classical Drude model, which is sufficient to prove the fundamental difference to a Corbino geometry. A Corbino disk is a two-terminal device, where one of the two contacts is entirely surrounded by a two-dimensional electron system (2DES). The second contact covers either the complete or part of the outer edge of the conductor. In this two-terminal device, the 2DES has no edge which connects contacts at different chemical potential. Consequently, there exists no Hall voltage and at constant applied voltage the electric field distribution remains constant, too, independent of the magnetic field. In a large perpendicular magnetic field, the Lorentz force then drives the electrons in closed trajectories perpendicular to the electric field along the equipotential lines which form closed loops around the inner contact. Correspondingly, the Drude model predicts the current to decrease with growing $B$ proportional to $\left[1+(\mu B)^2\right]^{-1}$.

\begin{figure*}[th]
\includegraphics[width=0.8\textwidth]{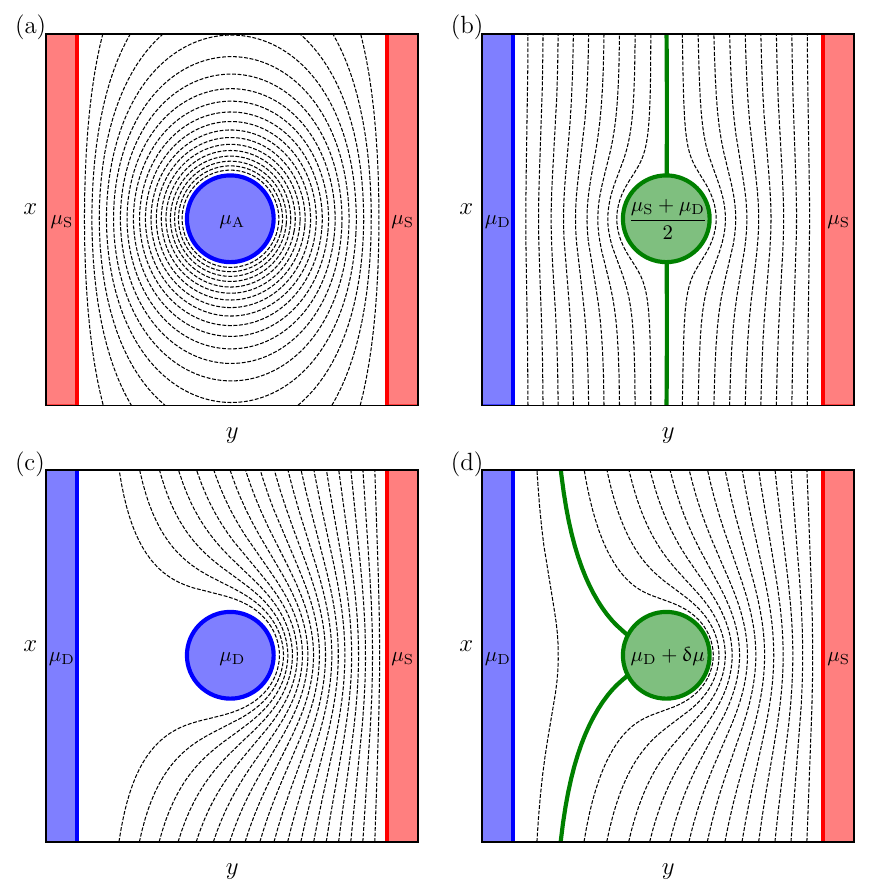}%
\caption{Sketches of a short section of a Hall bar including an inner contact (for simplicity with circular symmetry) for various chemical potentials assumed at the two edges and the inner contact. The equipotential lines of the Hall field (thin dashed lines) are numerically calculated for the two-dimensional geometry.
(a) Both edges have the identical chemical potential $\mu_\mathrm{S}$ (red) differing from that of the inner contact, $\mu_\mathrm{A}$ (blue). The equipotential lines encircle the inner contact resembling a Corbino disk.
(b) The two edges have different chemical potentials, $\mu_\mathrm{S}$ and $\mu_\mathrm{D}$, and the inner contact is balanced at $\mu_\mathrm{A}=(\mu_\mathrm{S}+\mu_\mathrm{D})/2$, as if it where electrically floating.The potential distribution is perturbed by the inner contact but the central equipotential line has the potential $\mu_\mathrm{A}$ and terminates at the inner contact (green line). For $\mu B\gg1$ current flowing along this line would connect with the inner contact.
(c) One edge has $\mu_\mathrm{S}$, the other edge and the inner contact have $\mu_\mathrm{D}$ as if they were electrically connected with zero resistance. In this extreme case no equipotential line terminates at the inner contact, which then were electrically isolated for $\mu B\gg1$.
(d) We assume a chemical potential around contact A of $\widetilde\mu_\mathrm{A}=\mu_\mathrm{D}+\delta\mu$ with $\delta\mu=(\mu_\mathrm{S}-\mu_\mathrm{D})/8$. This is the most realistic scenario. As $0<\delta\mu<|\mu_\mathrm{S}-\mu_\mathrm{D}|$, there exists always one equipotential line with the potential $\widetilde\mu_\mathrm{A}$, which terminates at the inner contact (green line). For $\mu B\gg1$ current flowing along this line would connect with the inner contact.
}
\label{fig:contactA}
\end{figure*}

Our three-terminal setup is a Hall bar with two outer contacts (S and D in Fig.~\ref{fig:sketch_suppl}), which are connected by edges of the 2DES. It would correspond to the Corbino geometry, if the two outer contacts (S and D) were forced to an identical chemical potential, while the inner contact would be at a different chemical potential. In Fig.\ \ref{fig:contactA}(a)
we present this Corbino equivalent. The dashed lines are numerically calculated equidistant equipotential lines, which are magnetic field independent. For simplicity, we assume a circular inner contact.

Our Hall measurements are fundamentally different from a Corbino scenario as there exists a potential drop between the outer contacts (S and D). We drive a constant current through the source contact (corresponding to a voltage applied to S), while the other two contacts (A and D) are connected to ground via ohmic resistances, $R_\mathrm{A}\simeq3.4\,\mathrm{k}\Omega$ and $R_\mathrm{D}\simeq2.4\,\mathrm{k}\Omega$, respectively. For $B>0.3\,$T most of the current flows between the outer contacts (S and D), while ---depending on the magnetic field--- a small percentage of it flows through the inner contact A (cf.\ Fig.~\ref{fig:sketch_suppl}). The fundamental difference from a Corbino device are the edges of the 2DES connecting current carrying contacts, because in a perpendicular magnetic field these edges will charge up. Whether current flows through the inner contact or not depends on how the corresponding Hall field, $\vec E_\mathrm{H}=\vec j \times \vec B / (e n_\mathrm{s})$, is modified by the presence of the inner contact.

To demonstrate the difference between our three-terminal setup and a Corbino device, we first consider the situation with the inner contact electrically floating ($R_\mathrm{A}\to\infty$). Assuming, that the contacts carrier density is much larger than that of the 2DES, it will perfectly shield the electric field as shown in Fig.\ \ref{fig:contactA}(b). The chemical potential of the floating contact will be half way between that of the edges, $\mu_\mathrm{A}=0.5\left(\mu_\mathrm{S}+\mu_\mathrm{D}\right)$. The central equipotential line terminates at the contact. Accordingly, a small part of the current flowing inside the 2DES along the equipotential lines flows through the floating contact A.

If the inner contact is connected to the electrical ground, its effect on the Hall field is different. The current flowing through an ohmic contact to ground is proportional to its chemical potential. In a two-terminal Hall bar, $\mu_\mathrm{S}=eR_\mathrm{S}I_\mathrm{S}$ and $\mu_\mathrm{D}=eR_\mathrm{D}I_\mathrm{S}$ are the chemical potentials along the two outer edges (assuming $I_\mathrm{A}=0$). This implies that the complete Hall potential, $eV_\mathrm{H}=\mu_\mathrm{S}-\mu_\mathrm{D}$, drops near the contacts \cite{Klass1991,Komiyama2006}. The details of how exactly the potential drops in a small transition region between the Hall bar and a contact, depend on the depletion of the 2DES near the contact. However, they are not important for the dynamics far away from the contacts. In contrast, if we are interested in how much current flows through the inner contact, these details become important, because what matters is the electric potential on the outside of this transition region. For instance, the chemical potential inside the ohmic contact is  $\mu_\mathrm{A}=eR_\mathrm{A}I_\mathrm{A}$, while the relevant potential outside of the transition region would be $\widetilde\mu_\mathrm{A}=e(\delta R+R_\mathrm{A})I_\mathrm{A}$ with $\delta R$ being a measure for the screening or coupling efficiency of the inner contact. As a result, the effective potential of contact A stabilizes between the two extreme cases, $\mu_\mathrm{D}<\widetilde\mu_\mathrm{A}\le0.5\left(\mu_\mathrm{S}+\mu_\mathrm{D}\right)$. The limit $\widetilde\mu_\mathrm{A}\to0.5\left(\mu_\mathrm{S}+\mu_\mathrm{D}\right)$ also corresponds to the floating contact with $R_\mathrm{A}\to\infty$ discussed above.

The limit $\widetilde\mu_\mathrm{A}\to\mu_\mathrm{D}$ corresponds to $\delta R\to0$ (with $R_\mathrm{A}\simeq R_\mathrm{D}$). We present the according potential distribution in Fig.\ \ref{fig:contactA}(c). For $\mu B\gg1$ all current would flow to drain and the inner contact would be electrically isolated.

In a realistic sample, the effective potential of contact A stabilizes between the two extreme cases discussed above. As a result, we find a situation such as shown in Fig.\ \ref{fig:contactA}(d), where we assumed $\widetilde\mu_\mathrm{A}=\mu_\mathrm{D}+(\mu_\mathrm{S}-\mu_\mathrm{D})/8$. In a qualitative picture, the current flowing for $\mu B\gg1$ along the equipotential line terminating at A then contributes to $I_\mathrm{A}$.

\begin{figure}[th]
\includegraphics[width=.85\columnwidth]{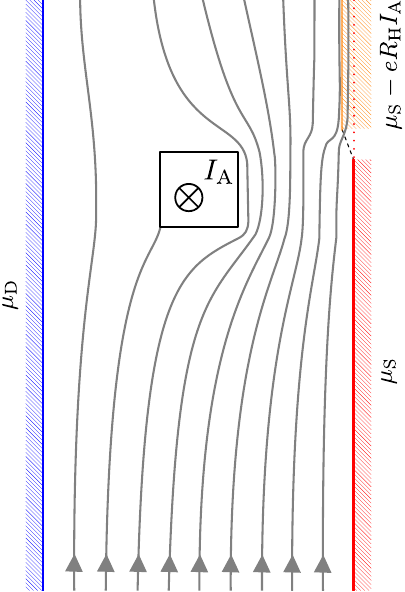}
\caption{Qualitative sketch accounting for the effect of $I_\mathrm{A}$ on the distribution of the Hall potential for $\mu B\gg1$. Upstream of the inner contact A the current density (arrows) is homogeneous. Current flows along the equipotential lines (solid lines) of the Hall field it generates. Downstream of the inner contact A, the sketched field distribution is out of equilibrium featuring an inhomogeneous current distribution.}
\label{fig:corrected_A}
\end{figure}

\begin{figure}[th]
\includegraphics[width=1\columnwidth]{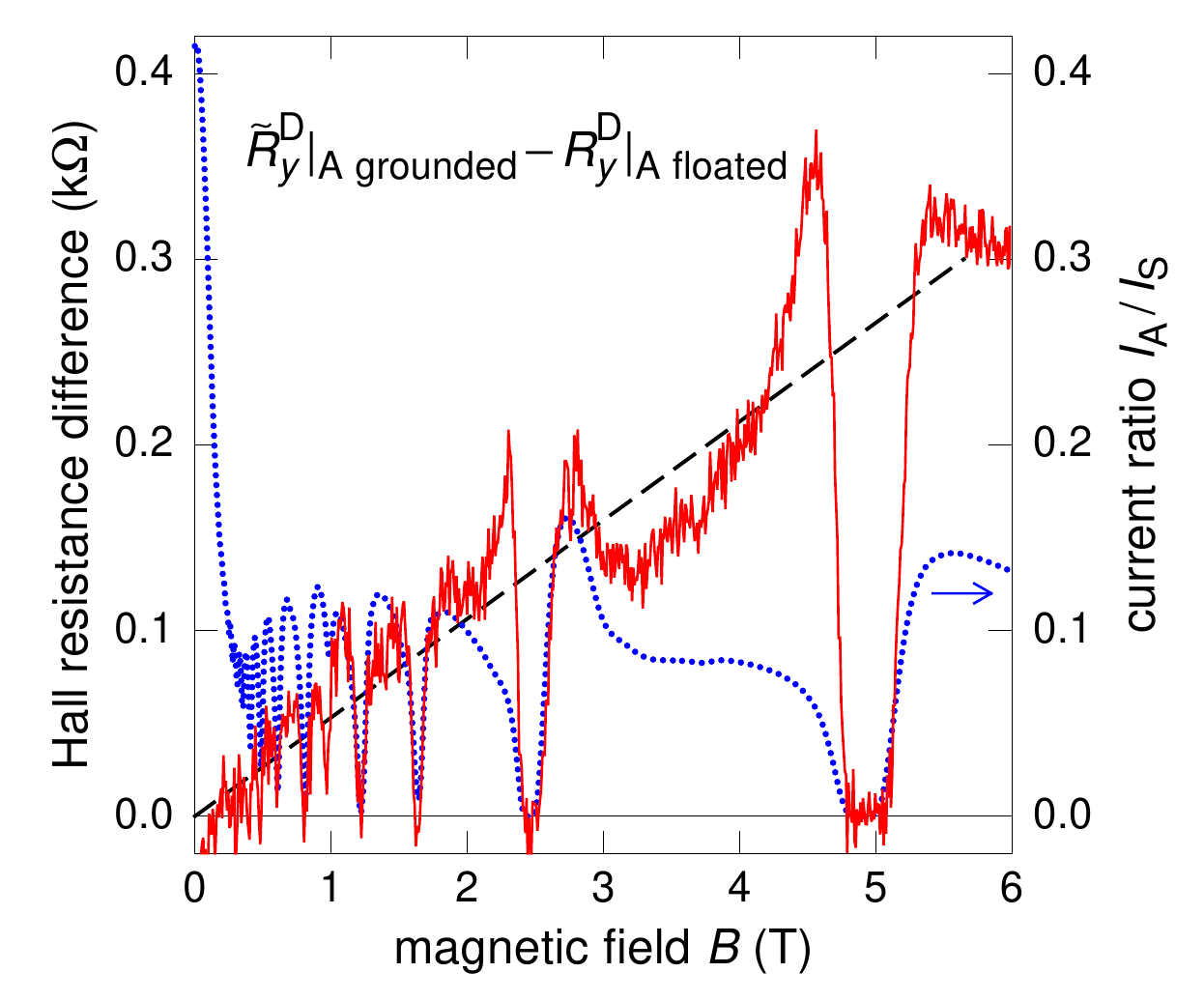}%
\caption{Difference between the Hall resistances with contact A
grounded versus A floated, $\widetilde{R}^\mathrm{D}_{y}|_\mathrm{A\ grounded}-R^\mathrm{D}_{y}|_\mathrm{A\ floated}$, as a function
of $B$ (solid red line) and the current $I_\mathrm{A}(B)$ flowing through A (dotted
line, right-hand-side axis). The black dashed line in the background
is proportional to $B$.}
\label{fig:res_diff}
\end{figure}

In the discussion so far we neglected the correction that $I_\mathrm{A}\ne0$ has on the distribution of the Hall potential, in particular downstream of A. In Fig.\ \ref{fig:corrected_A}
we qualitatively sketch the Hall potential distribution, still considering the diffusive regime. Upstream of A the current density and the Hall potential are distributed homogeneously across the Hall bar. For $\mu B\gg1$, the current (arrows) flows along the equipotential lines (solid lines). The current $I_\mathrm{A}$ branched off through the inner contact results in an inhomogeneous current density downstream of A: The reason is, that the current flowing into A is missing downstream of A. A consequence is a reduction of the chemical potential on the high potential edge of the Hall bar by $eR_\mathrm{H}I_\mathrm{A}$. Far away and downstream of A, we expect again a homogeneous current distribution. However, close to A the current density becomes inhomogeneous. It is reduced just beyond A and increased near the high potential edge. In particular, the enhanced current density near the high potential edge causes the Hall resistance $R_\mathrm{H}=B/(en_\mathrm{s})$ to be increased, because the carrier density $n_\mathrm{s}$ is smaller near the edges of the Hall bar. Like the Hall resistance, this increase is proportional to $B$. This effect explains the increased corrected Hall resistance  $\widetilde{R}^\mathrm{D}_{y}=\frac{V_2-V_3}{I_\mathrm{S}-I_\mathrm{A}}$ (for A grounded) compared to $R^\mathrm{D}_{y}=\frac{V_2-V_3}{I_\mathrm{S}}$ (for A floated) visible, e.g., in Fig.\ 3 of the main article away from the plateaus. In Fig.~\ref{fig:res_diff}
we present $\widetilde{R}^\mathrm{D}_{y}|_\mathrm{A grounded}-R^\mathrm{D}_{y}|_\mathrm{A floated}$ in order to visualize this excess Hall resistance. At its local maxima, $I_\mathrm{A}$ approximately increases proportional to $B$, which confirms our model.

\section{Current flowing through the inner contact versus breakdown of the QHE}\label{app:currdep}

The electrically induced breakdown of the QHE \cite{Kaya2000,Yu2018}, i.e., its dependence on the applied voltage was studied in Ref.\ \cite{Haremski2020} for various Hall bar widths. The authors reported that the breakdown depends on the width of the Hall bar at the high magnetic field ends of the plateaus while they found a much smaller width dependence at the low magnetic field ends of the plateaus. These findings support the screening theory predictions of bulk (edge) transport at the high (low) magnetic field ends of the plateaus: the observed breakdown behavior suggests that edge transport is independent of the Hall bar width while bulk transport depends on the width.

\begin{figure*}[th]
\includegraphics[width=0.85\columnwidth]{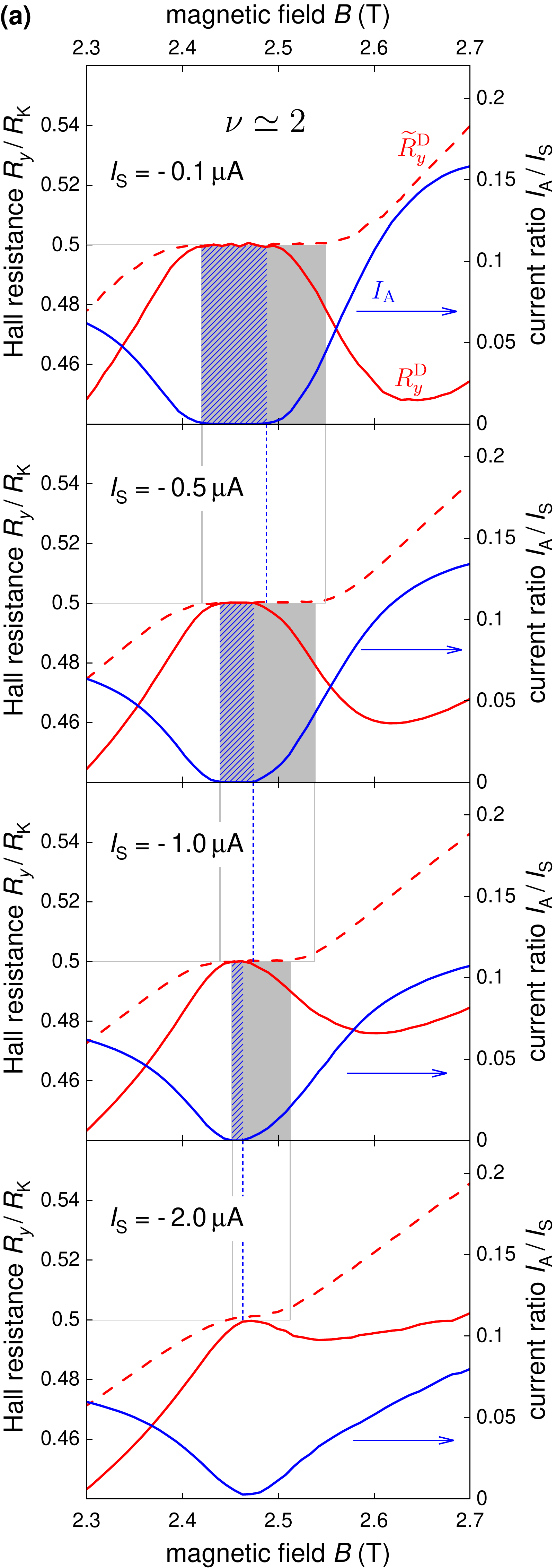}%
\hspace{20mm}
\includegraphics[width=0.85\columnwidth]{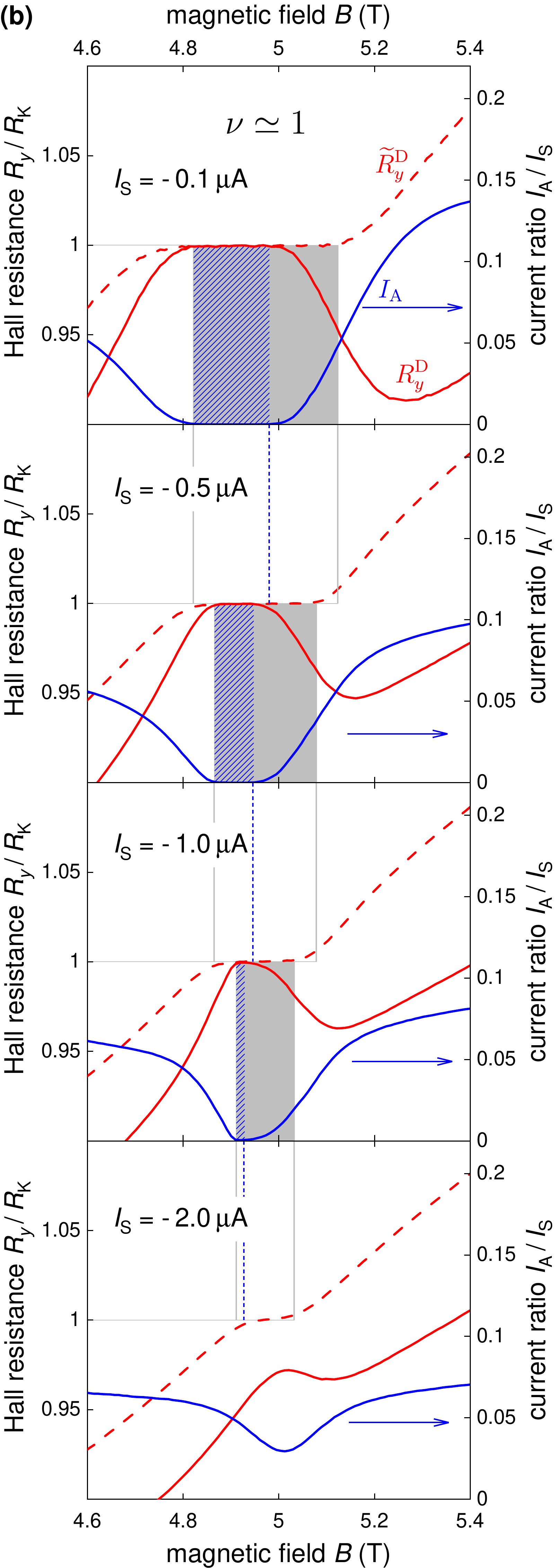}%
\caption{Hall resistances $R^\mathrm{D}_{y}(B)$ (red-solid) and $\widetilde{R}^\mathrm{D}_{y}(B)$ (red-dashed) as well as the current ratio $I_\mathrm{A}/I_\mathrm{S}(B)$ (blue solid line, right hand side axes) for filling factors $\nu\simeq2$ in (a) and $\nu\simeq1$ in (b). From top to bottom, the current $I_\mathrm{A}$ applied through the source contact is increased. The inner contact A is connected to ground. The gray areas indicate the plateau region with $\widetilde{R}^\mathrm{D}_{y}=R_\mathrm{K}/\nu$ and $\nu=1,2$. The blue patterned areas indicate magnetic field regions with $I_\mathrm{A}=0$ and $R^\mathrm{D}_{y}=R_\mathrm{K}/\nu$.
}
\label{fig:hall_cdep}
\end{figure*}

Here, we study the electrically induced breakdown as a function of the current applied through the source contact $I_\mathrm{S}$. At the same time, we explore the dependence of the current flowing through the grounded inner contact $I_\mathrm{A}$ on $I_\mathrm{S}$. In Fig.~\ref{fig:hall_cdep},
we present the Hall resistance $R^\mathrm{D}_{y}(B)=\frac{V_2-V_3}{I_\mathrm{S}}$, its corrected value $\widetilde{R}^\mathrm{D}_{y}(B)=\frac{V_2-V_3}{I_\mathrm{S}-I_\mathrm{A}}$ and the current through the inner contact $I_\mathrm{A}(B)$ for various values of the applied current $I_\mathrm{S}$, which is increased from top to bottom. We consider the filling factors near $\nu=2$ in Fig.~\ref{fig:hall_cdep}(a) versus $\nu=1$ in Fig.~\ref{fig:hall_cdep}(b). Areas shaded in gray approximately indicate the magnetic field regions of the plateau of the corrected Hall resistance, $\widetilde{R}^\mathrm{D}_{y}=R_\mathrm{K}/\nu$. The decrease of this gray area as $I_\mathrm{S}$ is increased indicates the electrically induced breakdown of the QHE. The breakdown happens on both ends of the plateau with subtle variations related with edge versus bulk transport at the two ends of the plateau \cite{Haremski2020}.

In Fig.~\ref{fig:hall_cdep} we also indicate as a blue stripe pattern the magnetic field regions of $I_\mathrm{A}=0$ (coinciding with $R^\mathrm{D}_{y}=R_\mathrm{K}/\nu$ and integer $\nu$). On the low magnetic field side the onset of $\widetilde{R}^\mathrm{D}_{y}<R_\mathrm{K}/\nu$ coincides with the onsets of $I_\mathrm{A}\ne0$ and $R^\mathrm{D}_{y}<R_\mathrm{K}/\nu$ (with integer $\nu=1,2$). The behavior is quite different at the high magnetic field side of the plateau. As discussed in the main article the onset of $I_\mathrm{A}\ne0$ with $R^\mathrm{D}_{y}<R_\mathrm{K}/\nu$ do not coincide with the onset of $\widetilde{R}^\mathrm{D}_{y}>R_\mathrm{K}/\nu$. The reason is the bulk current presented by $I_\mathrm{A}$, which flows through the Hall bar. As we increase $I_\mathrm{S}$ the onset of $I_\mathrm{A}\ne0$ shifts to smaller $B$ independently of and more rapidly than the breakdown induced decrease of the plateau width (gray areas). This finding indicates, that the magnetic field region of scattering-free bulk current increases with growing $I_\mathrm{S}$.

This behavior is indeed predicted by the screening theory: Starting from the low magnetic field side of the plateau we consider two ICSs, one along each edge. A higher current $I_\mathrm{S}$ increases the width of the  ICS near the high potential edge of the Hall bar as it has to carry more current (e.g., see Figure 9 in Ref.\ \cite{Gerhardts2008}. The wider ICS naturally fosters the transition from edge to bulk transport, such that $I_\mathrm{A}\ne0$ starts at smaller values of $B$.

In summary, our $I_\mathrm{S}$ dependent measurements qualitatively confirm the predictions of the screening theory for the current dependence of the transition from edge to bulk transport.

%